\documentclass[aps,nofootinbib,twocolumn]{revtex4}

\usepackage[latin1]{inputenc}
\usepackage{amsfonts}
\usepackage{hyperref}
\usepackage{amssymb}
\usepackage{amsmath}
\usepackage{graphicx}
\usepackage{epsfig}
\usepackage{bm}
\usepackage{ulem}
\usepackage{bbold}

\def\bra#1{\mathinner{\langle{#1}|}}
\def\ket#1{\mathinner{|{#1}\rangle}}
\def\mn#1{\langle #1 \rangle}
\def\prjct#1{\mathinner{|{#1}\rangle}\!\!\mathinner{\langle{#1}|}}

\def\text#1{\textrm{#1}}
\def\id{\mathbb{1}}

\begin{document}

\title{Cloning Entangled Qubits to Scales One Can See}
\author{Pavel Sekatski$^1$}
\email{Pavel.Sekatski@unige.ch}
\author{Bruno Sanguinetti$^1$}
\author{Enrico Pomarico$^1$}
\author{Nicolas Gisin$^1$}
\author{Christoph Simon$^2$}
\affiliation{$^1$ Group of Applied Physics, University of Geneva, 20~rue de l'Ecole-de-M\'edecine, CH-1211 Geneva 4, Switzerland
\\$^2$ Institute for Quantum Information Science and Department of Physics and Astronomy,
University of Calgary, Calgary T2N 1N4, Alberta, Canada}

\begin{abstract}
By amplifying photonic qubits it is possible to produce states that contain enough photons to be seen with a human eye, potentially bringing quantum effects to macroscopic scales\cite{Sekatski}. In this paper we theoretically study quantum states obtained by amplifying one side of an entangled photon pair with different types of optical cloning machines for photonic qubits. We propose a detection scheme that involves lossy threshold detectors (such as human eye) on the amplified side and conventional photon detectors on the other side. We show that correlations obtained with such coarse-grained measurements prove the entanglement of the initial photon pair and do not prove the entanglement of the amplified state. We emphasize the importance of the detection loophole in Bell violation experiments by giving a simple preparation technique for separable states that violate a Bell inequality without closing this loophole. Finally we analyze the genuine entanglement of the amplified states and its robustness to losses before, during and after amplification.
\end{abstract}

\maketitle

\section{Introduction}

The cloning of a photonic qubit is an intriguing subject, especially when the number of clones is so large that one can see them with the naked eye.
Such a macroscopic photonic state is worth analyzing, especially when the initial qubit is entangled with a twin photon~\cite{Sekatski} as illustrated in Fig.~1. This is the main purpose of this article, though still at the theoretical level.

Naively, one could think that if one has access to a number of clones, even of poor quality, then one could measure the state in which these clones are and infer the state of the initial qubit. This idea, applied to half of an EPR entangled pair of photons, is the basis of the faster-than-light flash telegraph proposed by Herbert in the 1980's~\cite{Herbert}. The no-cloning theorem was actually motivated by the impossibility of this flash telegraph~\cite{WoottersZurek,Diekes,Milonni}. Today, it is known that optimal quantum cloning has a fidelity which is precisely at the limit of  the flash telegraph: if one could clone qubits any better than allowed by quantum physics, then one could use this process, together with entanglement, to signal at arbitrarily high speeds~\cite{Gisin,BuzekGisinSimon}.

The reason why one can't extract more information from the many clones obtained by optimal quantum cloning than from the original qubit is that the many clones are in a complex entangled state~\cite{GisinMassar}. Indeed, would they be in a product state, then one could measure them individually and accumulate a lot of information from which the initial qubit state could indeed be inferred. But, actually, for any arbitrary number of clones, if optimal, their entangled state contains precisely the same information as the original qubit~\cite{GisinMassar,Acin}. Hence, a large number of clones constitute a sort of macroscopic "qubit". Note however, that this macroscopic state is not a real qubit: it lives in a Hilbert space of dimension much larger than two (the only exception is the optimal phase covariant cloner with precisely N output photons with zero loss). Nevertheless, we adopt the terminology  macroscopic "qubit" as a shorthand for the macroscopic quantum state produced by cloning of a single-photon qubit.

One of the fascinating aspect of a macroscopic photonic "qubit", is that one should be able to see it with the naked eye. What exactly can be inferred from such a direct observation of a "qubit" is the central issue of this paper. The eye is, to a good approximation, a 7-photon threshold detector preceded by about 90\% loss between the pupil and the retina~\cite{riekebaylor} (Nature doesn't always produce perfect optics~\cite{MantisSchrimps} !). In this article "to {\it see}" means to detect using two threshold detectors with postselection of the cases where one and only one detector went above the threshold.

\begin{figure}
\includegraphics[width=9cm]{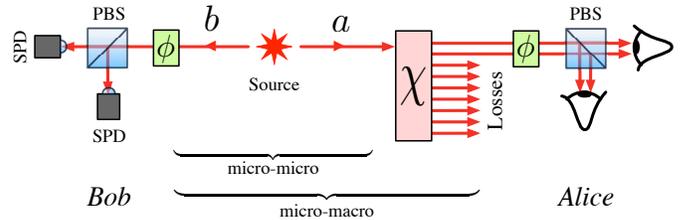}
\caption{ A source produces 2 entangled photons. The left one is measured using single-photon detectors (SPD). The right hand side one is first amplified ($\chi$) and the detection is made by 2 "eyes" modelled as losses and 2 threshold detectors.}
\label{fig:setup}
\end{figure}

This article addresses four main questions. Firstly, can one {\it see} the entanglement that exists between the two initial photons, i.e. can one {\it see} the micro-micro entanglement (see first line of Table 1)? By "{\it seeing entanglement}" we mean to infer it from the correlations between the two subsystems, where one is observed with lossy threshold detectors as shown in Fig.~\label{fig:setup}.
Next, is there, in principle, any entanglement between the macroscopic "qubit" and the twin photon (line 2 of Table 1)? Thirdly, can one see any micro-macro entanglement between the macroscopic "qubit" and the twin photon (line 3 of Table 1)? And, finally, does the {\it seen} correlation violates a Bell inequality with post-selection (last line of the Table)? \\

\footnotesize
\hspace{-20pt}
\begin{tabular}{|c||c|c|c|}
  \hline
               &                  &  &                    \\
type of cloner & universal &   phase covariant        &  measure-\&-prepare \\
                 &                &  & \\

\hline\hline
can one see    &                  &                 &                 \\
micro-micro    &    yes           &    yes          & yes \\
entanglement   &                  &                 &\\
\hline
preserves      & yes (scales as      & yes          &     no              \\
entanglement   & 1 / \#clones)     & (ideally 1 e-bit) &\\
\hline
can one see    &                  &                 &                     \\
micro-macro    &   no             &   no            & no \\
entanglement   &                  &                 &\\
\hline
can violate      &                 &                 &                    \\
Bell with        &yes&           yes   &yes \\
post-selection   &                  &                 &\\
\hline
\end{tabular}\\
{This table summarizes the main results of the paper. By {\it seeing} we mean "detectable with threshold detectors and selection of the events where one and only one detector went over the threshold". "Preserves entanglement" means that there is some micro-macro entanglement between the single-photon on the left and the amplified quantum state on the right hand side; however this can't be seen. For the last line, let us emphasize that Bell violation can never be {\it seen} because of the intrinsic inefficiency of threshold detectors. However, as indicated, the post-selected data can violate Bell inequality thanks to the detection loophole. For further explanation, see the main text.}\\
\normalsize

But let us first consider the different kinds of cloners and the very different ways of realizing optimal quantum cloning, i.e. the columns of Table 1. Universal cloning (first column), which clones all possible qubits with equal fidelity, can be realized with a simple optical amplifier, e.g. an erbium doped fiber amplifier~\cite{Fasel}. Universal cloning necessarily comes with additional quantum systems, called anti-clones; in the mentioned example the additional systems are the erbium ions. The clones and anti-clones (e.g. photons and ions) are entangled, hence when one considers only the clones, one expects that their entanglement with the twin photon is reduced, as mentioned in Table 1 and quantified in section \ref{entanglement}. Phase covariant cloning (second column), that best clones all qubits on a grand circle of the Poincar\'e sphere, can be realized by parametric amplification, as has been demonstrated for large numbers of clones by the Rome group~\cite{Rome, Rome2}, who did a brilliant work that stimulated ours.
%
%
If the pump of the amplifier is treated as a classical field
then it does not retain any information on the number of photons ($n$)  that have
been transferred into the amplified mode. The output is then a coherent sum
(superposition) of all possible outcomes with different $n$, consequently phase covariant cloning is a unitary transformation;
it thus preserves entanglement, see section \ref{entanglement}. For large numbers of clones, the optimal fidelity of universal and of phase-covariant cloning are 2/3 and 3/4, respectively~\cite{optFidelities}. However, if the only figure of merit is the fidelity of the clones, then there is a much simpler way of obtaining many clones: simply measure the initial qubit in an arbitrary basis - or for phase-covariant cloning in an arbitrary basis on the grand circle - and produce many qubits in the state corresponding to the measurement result. When averaged over the measurement basis, one thus obtains universal or phase covariant clones with the same fidelity as optimal quantum cloning~\cite{Acin}. But, obviously, this measure-\&-prepare cloner (third column of the Table) is not identical to the amplifier based cloner. Indeed, when applied to one photonic qubit entangled with a twin photon, the measure-\&-prepare cloner clearly breaks the entanglement. Interestingly, this is not the case for the amplifier based cloners, as already mentioned. However, we shall see that the correlations observed with threshold detectors are almost identical whether one uses the measure-\&-prepare cloner or an amplification based cloner, hence it is clearly impossible to {\it see} micro-macro entanglement.

Next, we come to the question whether one can {\it see} micro-micro entanglement. Entanglement is a concept defined within Hilbert space quantum physics: a quantum state is entangled iff it can't be written as a product, but only as a sum of product states. Moreover, it is well known that all entanglement witnesses, except Bell inequalities, require assumptions on the dimension of the Hilbert space
\footnote{
If observed correlations are local, in the sense that they violate no Bell inequality, then there exist an underlying local model, with a local variable $\lambda$, that reproduces these correlations. And any particular local configuration $\lambda$ can be implemented as a produced state $\ket{\lambda}$ of a quantum system if we allow the dimension of its Hilbert space be arbitrarily big (potentially reproducing all the classical physics).
}
. Consequently, the question {\it Can one see entanglement?}, implicitly assumes standard quantum physics and a given Hilbert space ~\cite{dimH}. In practice, if one considers the cloning machine as part of the measurement apparatus, then one studies the correlations between the two micro-micro photons. While, if one considers the cloning machine as part of the source, then one studies the correlation between the micro-macro photonic qubits. In the first case the Hilbert space dimension is clearly 2x2; this case can thus be analyzed as usually and the use of cloners and threshold detectors with finite efficiency is similar to the case in which equally inefficient single-photon detectors are used (important, however, is that the efficiency has to be independent of the measurement settings). Accordingly, it is possible to {\it see} micro-micro entanglement. However, there is one big difference between that case of single-photon detectors and that of cloner + threshold detectors that we like to emphasize. In the first case, the measurement setting is chosen before the signal is amplified to the classical level, while in the latter case amplification happens before the choice of the measurement setting.

In the micro-macro case, the dimension of the relevant Hilbert spaces is more tricky. It depends on the kind of cloner and on the presence of losses. Generally, if there are $N$ photons, the state is a superposition of symmetric states with $n$ photons polarized vertically and $N-n$ polarized horizontally with $n=0...N$, hence there are $N+1$ orthogonal states. In brief, in the micro-macro case, the Hilbert space dimension is much larger than 2 and detecting entanglement should be much harder as we confirm in section \ref{entanglement}. Actually, as already mentioned, it is impossible to detect micro-macro entanglement with threshold detectors, i.e. it is impossible to {\it see} micro-macro entanglement.

Nonlocality, that is the violation of a Bell inequality, is a concept that, contrary to entanglement, goes beyond quantum physics: correlations are nonlocal iff they can't be reproduced by local variables. Hence, the locality/nonlocality of {\it seen} correlations has nothing to do with the quantum stuff measured, it is only a characteristic of the classical data collected by measurements. According to quantum physics, nonlocal correlations can only be obtained when measuring entangled states. Hence, the observation of nonlocal correlation is a signature that the measured quantum state was entangled, irrespective of any Hilbert dimension criteria. From this and the conclusion of the previous paragraph one can already conclude that it is impossible to {\it see} nonlocal correlations. Moreover, since the study of nonlocality goes beyond quantum physics, one has to be careful with finite efficient detection: the nondetected events could open the possibility to describe the experiment by local variables. This is the infamous detection loophole. This loophole is often misconsidered as esoteric, but we shall see that if not taken into account one can easily come to wrong conclusions. Indeed, we shall see that if one merely ignores nondetected events, one can violate the CHSH-Bell inequality even with an entanglement breaking cloner and threshold detectors, see section V.

\normalsize

\section{General Scenario}\label{general scenarion}

In a previous article~\cite{Sekatski} we discussed the possibility of designing  quantum experiments using one's naked eyes as detectors. We showed that this can be done with a cloning machine, which amplifies the state to visible levels before the choice of measurement basis is made. The setup of the proposed Bell experiment is shown in Figure~(\ref{fig:setup}): first, an entangled photon pair singlet state $\ket{\psi^-}= \frac{1}{\sqrt{2}}(a^\dag b_\perp^\dag -a_\perp^\dag b^\dag)\ket{0}$ is produced by parametric down conversion, where $a^\dag, a_\perp^\dag (b^\dag, b_\perp^\dag )$ are the creation operators for two orthogonal polarizations in the spatial mode $a \,(b)$,. One photon is measured, in any desired basis, with standard single photon detectors, which for simplicity we assume to have negligible noise.
The other photon is amplified (cloned) and the two desired orthogonal polarization components are  measured with human eyes.
We consider a measurement to be conclusive if one and only one eye sees a signal.
This kind  of post-selection allows one to violate the CHSH Bell inequality using any of the 3 types of amplification presented in the introduction. In this article we answer the following questions: does the entanglement present in the initial state survive the amplification process? What can one deduce from the measurement statistics of this experiment?




\section{Types of Cloning Machines}\label{cloning}

In this section we review the 3 kinds of cloning machine considered in this article: the Universal, Phase Covariant and Measure and Prepare Cloning Machines. For each one we introduce the mathematics describing the amplification of an input qubit to a large number of output qubits.

\subsection{The Universal Cloning Machine}
A Universal Quantum Cloning Machine (UCM) is a device that clones all qubits on the Poincaré sphere equally well.
In the case of photonic polarization  qubits, such a device can be experimentally achieved by a type-II down conversion, seeded by the qubit (photon) to be cloned. The clones will be produced in the ``signal'' mode,  i.e. the mode excited by the input photon.
The photons in the idler mode, called the anticlones,  are disregarded.
In fact any phase invariant amplifier, such as an Erbium doped fibre amplifier (EDFA), can be used as a universal cloning machine.
It is to be noted that the transformation implemented by a universal cloning machine is not unitary, as the information carried by the anticlones is discarded.

The interaction Hamiltonian for an optimal universal quantum cloning machine can be written as \cite{OptClonLett}:
\begin{equation}\label{universal}
H = i \kappa (a^\dag_h c^\dag_v - a^\dag_v c^\dag_h ) + \textrm{h.c.}\,,
\end{equation}
where $a$ is the input mode to be cloned and $c$ is an auxiliary mode that is traced out, i.e. the anti-clone mode. This Hamiltonian has the singlet form, and is thus invariant over the Poincar\'e sphere (of course we can freely rotate the anti-clones because they are traced out). Hence, it is valid for any two orthogonal polarizations and we can write $a$ and $a_\perp$ instead of the arbitrary horizontal and vertical polarizations $a_h$ and $a_v$. The resulting propagator, $U_t=\exp(g (a^\dag c^\dag_\perp -a c_\perp))\exp(-g (a^\dag_\perp c^\dag -a_\perp c)=U U_\perp$, decomposes into a product of orthogonal polarization terms. In Appendix \ref{algebra} we show how the ``disentanglement  theorem'' (\ref{desentanglement}) can be used to express  $U$ in the following ordered form:
\begin{equation}
U = e^{T_g a^\dag c^\dag_\perp} C_g^{-(a^\dag a +c^\dag_\perp c_\perp+1)}e^{-T_g\, a c_\perp},
\end{equation}
where $T_g=\tanh(g)$ and $C_g= \cosh(g)$. The reason why such reordering formulas exist is that the set of operators $\{a^\dag b^\dag, a \,b, a^\dag a, b^\dag b\}$ is closed with respect to the commutator, see appendix for details. Initially the mode ${c_\perp}$ is taken to be in the vacuum state $\ket{0}$, i.e. an empty idler mode in a down conversion experiment, or a totally inverted atomic population in an EDFA. The propagator acting on the vacuum ${c_\perp}$-state is
\begin{equation}\label{amplichannel}
\textrm{C}_A=U \ket{0}_{c_\perp}= e^{T_g a^\dag c^\dag_\perp} C_g^{-a a^\dag }\ket{0}_{c_\perp}.
\end{equation}
A similar expression for $\textrm{C}_A^\perp$ can be obtained for the orthogonal mode $a_\perp$. If we put the two modes together the universal amplification is then given in the usual terms of the generalized evolution by the superoperator $\$_A(\rho)$ acting on the input state $\rho$:
\begin{equation}
\$_A(\rho)= tr_{c,c_\perp} \textrm{C}_A \textrm{C}_A^\perp \, \rho \, \textrm{C}_A^\dag \textrm{C}_A^{\perp\dag},
\end{equation}
where $tr_{c_\perp} \textrm{C}_A \, \rho \, \textrm{C}_A^\dag = \sum_i \textrm{C}_A^i \,\rho\,\textrm{C}_A^{i\dag}$ and $\textrm{C}_A^i =\frac{ \tanh^i(g) }{\sqrt{i!}}a^{\dag \,i} \cosh(g)^{-a a^\dag}$.
In order to find the fidelity of this cloning process we need to find the evolution of the photon number operator $a^\dag a\rightarrow U^\dag a^\dag a U$.

In appendix \ref{Fock space projections} we show that in general it is possible to obtain the mean values of any power of the photon number operator, together with any higher order correlation function, from the knowledge of the mean value of the characteristic function $N(k)=\text{tr} \,e^{k a^\dag a} \rho$. The reordering relations of appendix \ref{algebra}
can be used to show that the operator $e^{k a^\dag a}$ is transformed by the universal cloner to:
\begin{equation}\label{UCchar}
\textrm{C}_A^\dag e^{k a^\dag a}\textrm{C}_A = e^{k a^\dag a}(C_g^2-S_g^2 e^{k})^{-a^\dag a-1}.
\end{equation}

The result (\ref{Nfunction}) implies that from a single photon input state $\ket{1}$ a universal cloner will yield a number of photons $n_1 = \partial_k \bra{1}\textrm{C}_A^\dag e^{k a^\dag a}\textrm{C}_A \ket{1}|_{k=0}$ in the same polarization mode as this input, while the orthogonal polarization mode will be populated by spontaneous emission $n_0 = \partial_k \bra{0}\textrm{C}_A^\dag e^{k a^\dag a}\textrm{C}_A \ket{0}|_{k=0}$, corresponding to the amplification of the vacuum state. Using equation (\ref{UCchar}) we obtain the well known result $n_1 = 2 \sinh^2(g)+1$ and $n_0= \sinh^2(g)$. The fidelity of the cloning process is defined as the ratio of the photon number in the input mode to the total photon number after the amplification. The optimal fidelity for large gains is then $\frac{2}{3}$. In appendix \ref{fibers} we show that the above treatment is also valid for universal cloners based on doped fiber amplification.

\subsection{The Phase Covariant Cloning Machine}
Phase covariant amplification is the cloning procedure that produces equally good and optimal clones of all states lying on a great circle of the Poincaré sphere, e.g. the equator. For photonic polarization qubits it can be achieved~\cite{Rome,Rome2} in degenerate collinear parametric type-II down conversion. Contrary to the case of universal cloning described above, here all the photons produced are kept in the output state. Hence, in the ideal lossless case, the transformation is unitary.

The Hamiltonian of the process reads:
\begin{equation}\label{PCCh-v}
H = i \chi a_h^\dag a_v^\dag + h.c.\,,
\end{equation}
with $\chi$ proportional to the non-linear $\chi^{(2)}$ susceptibility and to the power of the pump, which is considered to be classical. Any choice of basis \{$a_\phi$, $a_\phi^\perp $\} with \{$a_h=\frac{e^{i\phi}}{\sqrt{2}}(a_\phi+ia_{\phi\perp})$, $a_v=\frac{e^{-i\phi}}{\sqrt{2}}(a_\phi- i a_{\phi\perp})$\} defines the same great circle of the Poincaré sphere that we henceforth call the {\it equator}. One can rewrite the Hamiltonian in a form explicitly invariant on this equator:
\begin{equation}
H = \frac{i \chi}{2}({a_\phi^\dag}^2+{a_{\phi \perp}^\dag}^2)+h.c.
\end{equation}
Henceforth, we omit the $\phi$ index of the particular basis, writing $a$ and $a^\dag$ for $a_\phi$ and $a_\phi^\dag$, respectively.
The evolution propagator, $U_t=exp(-i t H)$, decouples on the two orthogonal polarization modes $U_t= U \otimes U_\perp$ where
\begin{equation}
U = e^{\frac{g}{2}({a^\dag}^2-a^2)}= e^{\frac{1}{2}T_g{a^\dag}^2}C_g^{-a^\dag a-\frac{1}{2} }e^{-\frac{1}{2} T_g a^2}
\end{equation}
with $C_g = \cosh(g)$, $T_g = \tanh(g)$ and $g = t \chi$. To derive the second equality we again used  the disentanglement theorem~(\ref{desentanglement}). The propagator for the orthogonal mode $a_\perp$ has the same form.
In an analogous way to the above treatment of the universal cloner, the reordering techniques of appendix~\ref{algebra} can be used to show that
the operator $e^{k a^\dag a}$ is transformed by the amplification process to $U^\dag e^{k a^\dag a} U$ :
\begin{align}\label{Pccgen}
U^\dag e^{k a^\dag a} U & = e^{-\frac{k}{2}}e^{\frac{T_g}{2}\left(\frac{e^{2k}}{C_g^2- S_g^2 e^{2k}}-1\right){a^\dag}^2} \times \nonumber\\
&\times \left(e^{-k}C_g^2-e^k S_g^2\right)^{-(a^\dag a+\frac{1}{2})} e^{\frac{T_g}{2}\left(\frac{e^{2k}}{C_g^2- S_g^2 e^{2k}}-1\right)a^2}.
\end{align}
From this expression we obtain photon numbers of $n_1= 3 \,\sinh^2[g] + 1$ and $n_0 = \sinh^2[g] $ for input states $\ket{1}$ and $\ket{0}$ respectively. The fidelity of cloning a qubit on the equator is then $F=\frac{3}{4}$ in the limit of large gains.

We emphasize that the phase covariant cloning is indeed bad in the $h$-$v$ basis: from the form of the hamiltonian (\ref{PCCh-v}) it is clear that the cloner puts the same number of photons in $h$ and $v$ modes and the initial difference in their populations is conserved. So the states $U_t\ket{1_h}$ and $U_t\ket{1_v}$ are different by only one photon in each mode and of course, when the total population becomes large, such a difference is unobservable with any realistic measurement precision.

\subsection{The Measure \& Prepare Cloning Machine}
\begin{figure}[htbp]
\begin{center}
\includegraphics[width=8.5cm]{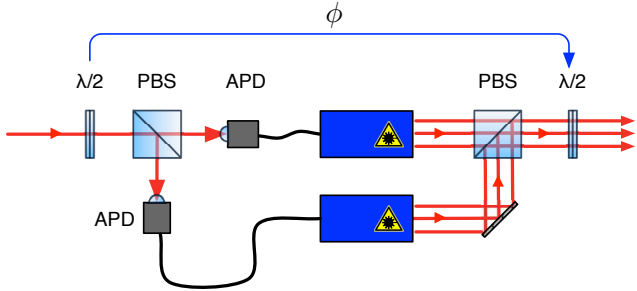}
\caption{A possible implementation of a Measure \& Prepare Cloning Machine: the two waveplates have the same angle.}
\label{fig:HCM}
\end{center}
\end{figure}

The operation of both the Universal and Phase Covariant cloning machines relies on quantum processes. It is interesting to see how these cloners compare to a more ``classical'' approach: one can simply measure the input state, in an arbitrary basis, and produce a stronger output according to the measurement result. An illustration of how such a device could be implemented is shown in figure~\ref{fig:HCM}.

The symmetry of this cloning machine is determined by the set of measurement bases amongst which a random basis is chosen before each measurement. For example, in the setup of figure~\ref{fig:HCM}, the use of half-wave plates would allow for a random choice of basis in a great circle of the Poincar\'e sphere. Restricting our discussion to the case of single-photon polarization qubits and large gains allows one to draw an analogy between this cloner and a phase-covariant cloner, as we show below~\cite{Acin}.

The single-photon is measured in a basis $\{a_\phi, a_{\phi\perp}\}$, with randomly chosen $\phi$. Depending on the measurement result, this cloner will prepare an output  coherent state in one of the two polarizations: $\ket{\alpha_\phi}$ or $\ket{a_{\phi\perp}}$.
Accordingly, for any input state $\rho_0$, the output state reads:
\begin{equation}\label{mp}
\rho = 2\int_0^{2\pi} \frac{d\phi}{2\pi}\bra{1_\phi}\rho_0\ket{1_\phi}\ket{\alpha_\phi}\!\!\bra{\alpha_\phi},
\end{equation}
where the factor 2 is due to the fact that for each measurement basis there are two possible outcomes that give twice the same term after the integration.

To prove the optimality of this cloning procedure, consider an input state $\ket{1}=a^\dag\ket{0}$ on the equator:
\[
\rho = \int \frac{d\phi}{\pi}c_\phi^2 \ket{c_\phi \alpha_a }\ket{s_\phi \alpha_{a\perp}}\!\!\bra{c_\phi \alpha_a }\bra{s_\phi \alpha_{a\perp}},
\]
where $(c_\phi,s_\phi)$ represent $\cos(\phi)$ and $\sin(\phi)$ respectively.
The output photon numbers are then given by:
\begin{alignat}{4}
n_a &= 2 |\alpha|^2\int\frac{d\phi}{2\pi}c^4_{\phi} &= \frac{3}{4}|\alpha|^2,\,\\
n_{a\perp} &= 2|\alpha|^2\int\frac{d\phi}{2\pi}c^2_{\phi}s^2_{\phi} &= \frac{1}{4}|\alpha|^2.
\end{alignat}
It is interesting to note that the fidelity of this process is $\frac{3}{4}$, the same as for the optimal phase covariant cloner, in the limit of large gains. Nevertheless, the output states of the two different machines are  very different: whilst the phase covariant cloner performs a unitary transformation, and therefore preserves entanglement, the Measure \& Prepare cloner trivially breaks the entanglement by projecting the state on a specific basis before amplification.

Just as the phase covariant cloner, the {\it measure \& prepare} cloner presented above is bad for the $h$-$v$ basis. Actually it is even worse: the input states $\ket{1_h}$ and $\ket{1_v}$ are both sent to the same output $\rho_{hv}=\int \frac{d\phi}{2\pi} \ket{\alpha_{\phi} }\!\!\bra{\alpha_{\phi} }$.

A {\it measure \& prepare} cloner with the same symmetry and fidelity as the universal cloner can be defined in a similar way, by choosing the measurement basis amongst the full set of points on the Poincar\'e sphere. The fact that we choose coherent states as output of the {\it measure \& prepare} cloner is only suggested by the simplicity of a potential realization, nothing prevents one to design such a cloner preparing states with any other photon number distribution. For example, with thermal states it gives a photon number distribution more similar to that of the state produced by the universal and phase covariant cloners.

\section{Detection of a Cloned State}\label{detection}
In order to evaluate what can be inferred from the detection of an amplified state, it is important to have a model of the available detector. In our case, we want to know how this state would be detected by the human eye: a lossy threshold detector, which we model as described below.

\subsection{Modelling Losses}
The simplest model of a detector with limited efficiency is a POVM, that is a combination of a loss mechanism and a Von Neuman projector on a part of Hilbert space. We describe the transmission loss by a beam splitter coupling an optical mode $a \to a \sqrt{\eta} + e \sqrt{1-\eta}$ to an inaccessible environment mode $e$, that is initially empty ($\hbar \omega\gg kT$ for visible wavelength at room temperature). We can express the loss channel $C_\eta$ as follows:
\begin{equation}
\textrm{C}_\eta= e^{\gamma( a^\dag e -  a \,e^\dag)} \ket{0}_e=e^{tan\,\gamma \,a e^\dag}cos(\gamma)^{a^\dag a} \ket{0}_e
\end{equation}
with $ \eta = \cos^2( \gamma)$, where we used (\ref{desentanglement}) for the second equality. An initial one mode state $\rho_a$ after a lossy transmission becomes $\rho_a' = \textrm{tr}_e \textrm{C}_\eta \rho_a \textrm{C}_\eta^\dag = \sum_{i\geq 0}^\infty \textrm{C}_\eta ^i \rho_a {\textrm{C}_\eta ^i}^\dag$
with $\textrm{C}_\eta^i =  \frac{tan^i(\gamma)}{\sqrt{i!}}a^i \cos(\gamma)^{a^\dag a}$. Of course a polarization encoded qubit undergoes transmission loss in both of the basis modes $a$ and $a_\perp$. If isotropic, such a loss channel is given by the product $\textrm{C}_\eta \textrm{C}_\eta^\perp$ with the same transmission $\eta$. Its action on an input state $\rho$ is given by the superoperator
\begin{equation}
\$_L(\rho) = tr_{e,e_\perp} \textrm{C}_\eta \textrm{C}_\eta^\perp \rho \textrm{C}_\eta^\dag \textrm{C}_\eta^{\perp\dag}.
\end{equation}
One can verify that this expression is indeed basis independent. The isotropic loss $\$_L$ and the universal amplification $\$_A$ are the two simplest processes with universal symmetry.

\subsection{Threshold Detectors}

 Let us consider a usual non-photon-number resolving detector with efficiency $\eta$. The probability of detection of the Fock state $\ket{n}$ is given by
\begin{equation}
\bra{n}\hat{P}_s\ket{n}= 1 - (1-\eta)^n.
\end{equation}
Such an operator $\hat{P}_s$ (the underscript $s$ stays for "see") is given by $\textrm{C}_\eta ^\dag\hat{I}_s\textrm{C}_\eta$, with $\hat{I}_s= \id-\ket{0}\!\!\bra{0}$ an ideal detector -  Von-Neuman projector and $\textrm{C}_\eta$ is the loss channel described above (it is possible to add a small universal cloner $\textrm{C}_A^\dag \hat{P}_s \textrm{C}_A$ to model the dark counts in the detector). It is natural to introduce a more general family - the threshold detectors, analogous to the ideal single-photon detector. An ideal threshold detector always "sees" the Fock states with the photon number $n$, if $n$ is bigger then the threshold $\theta$, and never "sees" Fock states with smaller photon number. In a realistic situation such an ideal detector is preceded by loss, giving the operator
\begin{equation}\label{treshold}
\hat{P_s^\theta} = \textrm{C}_\eta^\dag(\id -  \sum_{n\geq0}^\theta \ket{n}\!\!\bra{n} ) \textrm{C}_\eta =\id -\hat{P_{ns}^\theta}.
\end{equation}
In appendix \ref{Fock space projections} we show that for any state $\rho$ a projection on any Fock state can be recovered from the scalar function $\Pi(z)$ which, if we introduce loss and use rearrangement formulas of the appendix \ref{algebra}, becomes
\begin{equation}
\Pi(z)= \textrm{tr} \rho \,\textrm{C}_\eta^\dag\, z^{a^\dag a} \textrm{C}_\eta  = \textrm{tr}\, \rho (\eta \, z + 1-\eta)^{a^\dag a}.
\end{equation}
To make the formulas shorter we introduce the scalar function $\pi_\eta(z)=(\eta \, z + 1-\eta)$. A situation where a detector with an efficiency $\eta_d$ is preceded with a transmission channel with $ t = \eta_T$ is equivalent to a detector with an effective efficiency of $\eta = \eta_d \eta_T$.

\subsection{Detecting with the Human Eye}
The human eye is a surprisingly efficient detector, with virtually no dark counts. At low light intensities rod cells are responsible for photon detection. These cells are densely packed on the retina ($\sim10^5/\text{mm}^2$), where threshold logic is also present to provide noise reduction~\cite{Barlow}.
The  human eye can then be modeled as a lossy threshold detector (\ref{treshold}) with ``efficiency'' $\eta \approx 7\%$ and a threshold  $\theta\approx 7$) \cite{Eye}.
The amplified singlet pair contains enough photons to imagine Bell type experiments with naked human eyes. In the following section we will describe in more detail how the eye can be combined with simple logic to produce a ``macro-qubit'' analyzer.

\subsection{The ``macro-qubit" analyzer}

\begin{figure}[htbp]
\begin{center}
\includegraphics[width=8.5cm]{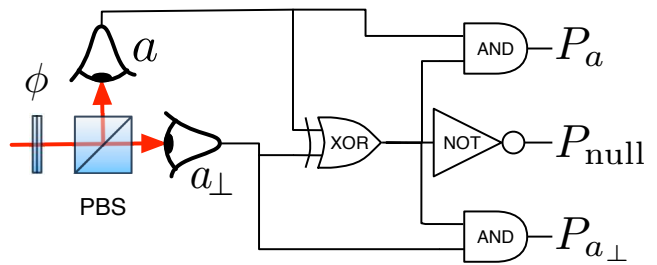}
\caption{Macro qubit analizer, consisting of a PBS followed by two lossy threshold detectors and  simple logic selecting conclusive events. The measurement basis can be chosen by applying a rotation ``$\phi$'' before the PBS. }
\label{fig:macro-qubit-analizer}
\end{center}
\end{figure}

Now we describe the full detection scheme on the amplified side (Fig.~\ref{fig:macro-qubit-analizer}). First the photons pass through a variable linear optical element that stands for the choice of the measurement basis among some variants (i.e. the setting of the measurement), then the two orthogonal polarization modes ($a$ and $a_\perp$) are separated on a PBS and sent to a corresponding threshold detector. We distinguish three possible outcomes. Firstly there are the two conclusive events (``$P_{a}$'' and ``$P_{a_\perp}$'') when only one of the two detectors sees. In these cases it is intuitively clear that statistically in the mode that triggered the threshold there were more photons than in the one that did not. Secondly there is the inconclusive event ``$P_\text{null}$'' that occurs when both detectors see or both do not. These three outcomes form a POVM, because they describe all the possible events. More formally we can write the corresponding operators as:
\begin{align}\label{POVM1}
\hat{P}_a &= \hat{P}_s^\theta(a) \hat P_{ns}^\theta(a_\perp)
\nonumber
\\
\hat{P}_{a_\perp} &= \hat P_{ns}^\theta(a) \hat P_s^\theta(a_\perp)
\\
\hspace{-30 pt}\hat{P}_\text{null}&=\id-\hat{P}_{a}-\hat{P}_{a_\perp}
\nonumber
\end{align}
According to the results shown in appendix 2, from the knowledge of the mean value $\mn{\pi_\eta(z)^{a^\dag a}\pi_\eta(z_\perp)^{a^\dag_\perp a_\perp}} = f(z,z_\perp)$ on a given quantum state we can directly infer the mean values of $\hat{P}_a$, $\hat{P}_{a_\perp}$ and $\hat P_\text{null}$ on the same state:
\begin{align}\label{generating}
\mn{\hat{P}_a} &= \frac{\partial_{z_\perp}^\theta}{\theta!}  \frac{f(1,z_\perp)}{1-z_\perp}|_0 -
\frac{\partial_{z_\perp}^\theta \partial_{z}^\theta}{\theta!^2}  \frac{f(z,z_\perp)}{(1-z_\perp)(1-z)}|_0
\nonumber\\
\mn{\hat{P}_{a_\perp}} &= \frac{\partial_{z}^\theta}{\theta!}  \frac{f(z,1)}{1-z}|_0 -
\frac{\partial_{z_\perp}^\theta \partial_{z}^\theta}{\theta!^2}  \frac{f(z,z_\perp)}{(1-z_\perp)(1-z)}|_0
\\
\mn{\hat{P}_\text{null}}&=1-\mn{\hat{P}_a}-\mn{\hat{P}_{a_\perp}}
\nonumber
\end{align}

\section{Inferring Micro-micro entanglement}\label{micro-micro}
In this section we consider the cloning machine as part of Bob's measurement apparatus and ask ourselves what can be concluded from the {\it seen} quantum correlation. More specifically, the question is whether one can conclude anything about the micro-micro quantum state produced by the source (see Fig.~1). Recall that Alice's measurement apparatus is conventional, while Bob's apparatus first uses one of the amplifier presented in section~\ref{cloning}, and then the ``macro-qubit'' analyzer presented above (Fig.~\ref{fig:macro-qubit-analizer}) to measure the amplified state.
The two measurement apparatuses look very different. However, on an abstract level, they are similar: both measure the polarization state of single-photons, both accept a polarization basis as measurement setting and both either produce a definite result or, with some probability, produce no result at all. The similarity goes on further: the probability that a measurement fails to produce an outcome is independent of the measurement setting, as proven below. Consequently, if both Alice and Bob's apparatuses are inserted in black boxes, then, from the outside, one can't distinguish them. The question of micro-micro entanglement between the two photons produced by the source is thus not different in our case as it is in usual, more common, measurement configuration, in particular in the configuration where Bob uses a measurement apparatus similar to Alice's one (with an efficiency that matches that of his ``measurement apparatus with human eyes'').

Accordingly, all the usual machinery of entanglement witnesses also apply to our configuration. In particular, one can {\it see} entanglement, that is one can conclude that the 2-photon source produces entanglement from the correlation observed with the naked eye, as one could conclude using standard PBS followed by single-photon detectors. In both cases one admits that detectors have a finite efficiency, independent of the measurement settings, and that the input state is well described as a pair of qubits, i.e. as a quantum state in the $2\otimes2$ Hilbert space.

We conclude this section emphasizing that the analysis would be completely different if we would not assume the cloning machine to be part of the measurement apparatus, as then one would be looking at possible entanglement between Alice's single-photon and Bob's large multi-photon quantum state, as discussed in the next section. The analysis would also be deeply different if one were looking for non-local correlations instead of mere entanglement. Indeed, as already mentioned in the introduction, entanglement is a quantum concept and can thus be analyzed within the usual Hilbert space machinery, accepting that measurement apparatuses sometimes fail to produce results (e.g. because of inefficient detectors), while the concept of nonlocal correlations goes beyond quantum physics. Hence, for the latter, no-result events can't be neglected, as briefly analyzed in the end of the section \ref{micro-micro}.F.

\subsection{Cloning as part of the detection apparatus}
Instead of considering the evolution of the input state through cloning we can see the cloning as a part of the detection process, i.e. as acting on the measurement apparatus. Of course from this perspective we completely disregard micro-macro entanglement, however it is much better fitted for analyzing the micro-micro entanglement aspect. All the complicated process of amplification and detection now acts on the small Hilbert space of the input state, i.e. the state of Alice's part of the system after an ideal detection of Bob's photon. This state lives in the space composed of a qubit component and a vacuum component, which accounts for a possible loss of Alice's photon. More formally the probability of a detection outcome $\hat{D}$ after the generalized evolution $\sum_i W_i \rho W_i^\dag$ of the input state $\rho$ is:
\begin{equation}
\text{tr}\;\hat{D}\sum_i W_i \rho W_i^\dag = \text{tr}\;\rho \sum_i W_i^\dag \hat{D} W_i  = \text{tr}\;\rho \hat{d},
\end{equation}
where $\hat{d} = \sum_i W_i^\dag \hat{D} W_i|_{H_{in}}$ is the operator acting on the Hilbert space of the input state and which  describes the same measurement outcome as $\hat D$.\footnote{Making the unitary evolution act on the operators $U^\dag \hat O U$, instead of states $U \rho U^\dag$, is called the Heisenberg picture as opposed to the Schrodinger picture. However, when dealing with the generalized evolution $\sum_i W_i \rho W_i^\dag$, one should take care when going to the Heisenberg picture, because if we naively write $\sum_i W_i^\dag \hat{O} W_i$ we implicitly trace out the auxiliary modes and in consequence such a transformation is not canonical in general. To do the things properly one has to introduce the Langevin noise operators in the operator transformation rules. However, if we are not doing any algebra with the transformed operators like in the case when the only quantity of interest is $tr \sum_i W_i \rho W_i^\dag \hat{O}$, it is unnecessary and we can stick with  $\sum_i W_i^\dag \hat{O} W_i$}
To make the restriction to the input Hilbert space explicit we can introduce twice its completeness relation $\ket{+}\!\!\bra{+}+\ket{-}\!\!\bra{-}+\ket{0}\!\!\bra{0} = \id$, with $\ket{+}$ and $\ket{-}$ staying for $a^\dag\ket{0}$ and $a_\perp^\dag\ket{0}$ - the two orthogonal single photon states in a certain basis. This basis choice allows us to rewrite $\hat{d}$ in its hermitian matrix representation:
\begin{equation}
\hat{d}=
\left (
\begin{array}{ccc}
d_{++} & d_{+-} & d_{+0} \\
d_{-+} & d_{--} & d_{-0} \\
d_{0+}&d_{0-}& d_{00}
\end{array}
\right )
\hspace{10 pt} \textrm{with}
\hspace{10 pt} d_{kl}= \bra{k}\hat{d}\ket{l}.
\end{equation}
In the micro-micro perspective any measurement outcome is fully described by these 9 real-value parameters, and the POVM in (\ref{POVM1}) is given by the set of three $3\times 3$ hermitian matrices $\{\hat{p}_a, \hat{p}_{a_\perp}, \hat{p}_\text{null}\}$ that sum to $\id$.

Let us illustrate how we calculate these operators: we showed how the mean value of the operators $\hat{P}_a$ and $\hat{P}_{a_\perp}$ can be obtained from that of $\pi_\eta(z)^{a^\dag a}\pi_\eta(z_\perp)^{a^\dag_\perp a_\perp}$ (section IV.D), consequently  the first step is to apply the amplification, described by the generalized evolution terms $W_i$, on $\pi_\eta(z)^{a^\dag a}\pi_\eta(z_\perp)^{a^\dag_\perp a_\perp}$:
\begin{equation}
\tilde{\Pi}(\eta, z, z_\perp...) =\sum_i W_i \pi_\eta(z)^{a^\dag a}\pi_\eta(z_\perp)^{a^\dag_\perp a_\perp}  W_i^\dag.
\end{equation}
The three points are here because the evolution can depend on various parameters such as gain and internal losses. We then explicitly restrict the operator $\tilde{\Pi}$ on the input Hilbert space by finding the nine scalar functions $\tilde{\Pi}_{ij}=\bra{i}\tilde{\Pi}\ket{j} $ with $\ket{i},\ket{j}$ belonging to the set of basis vectors of the input Hilbert space $\{\ket{+},\ket{-},\ket{0}\}$. Finally, by applying one of the relations from (\ref{generating}) to these functions $\tilde{\Pi}_{ij}$, we find the nine numbers that represent the corresponding operators of $\{\hat{p}_a, \hat{p}_{a_\perp}, \hat{p}_\text{null}\}$.

The contribution to the correlation visibility that we obtain from a measurement in the setting $\{a,a_\perp \}$ and $\{b,b_\perp \}$ is by definition $\mn{(\hat{p}_a-\hat{p}_{a_\perp})(\ket{1_b}\!\!\bra{1_b}-\ket{1_{b_\perp}}\!\!\bra{1_{b_\perp}})}$. After post-selection the observed visibility becomes:
\begin{equation}\label{vispost}
V_{a,a_\perp}=\frac{\mn{(\hat{p}_a-\hat{p}_{a_\perp})(\ket{1_b}\!\!\bra{1_b}-\ket{1_{b_\perp}}\!\!\bra{1_{b_\perp}})}}
{\mn{\hat{p}_a+\hat{p}_{a_\perp}  }}.
\end{equation}
For the singlet state this mean value is:
\begin{equation}\label{singletVis}
V_{a,a_\perp} = \frac{(\hat{p}_a)_{++}+(\hat{p}_{a_\perp})_{--}-(\hat{p}_a)_{--}-(\hat{p}_{a_\perp})_{++}}
{(\hat{p}_a)_{++}+(\hat{p}_{a_\perp})_{--}+(\hat{p}_a)_{--}+(\hat{p}_{a_\perp})_{++}}.
\end{equation}

Above we have shown how the correlation visibility for a particular amplification protocol can be evaluated. Next we will derive a bound on this correlation visibility that implies the presence of entanglement in the initial micro-micro state. To do this, in the following section we will introduce an entanglement witness.


\subsection{Entanglement witness}

The only entanglement witness that we will use in this paper is the one introduced in \cite{Witness}. It has the following form:
\begin{equation}\label{witness}
W = |\mn{\vec{J_A}\cdot \vec{J_B}}|-\mn{N_A N_B},
\end{equation}
where $\vec J_A$ and $\vec J_B$ denote Stokes vectors (total polarization) on parts $A$ and $B$, while $N_A$ and $N_B$ are the total number of photons for $A$ and $B$.
For any separable state $\sum_i p_i \rho_i \otimes \sigma_i$ we can bound $|\mn{\vec{J_A}\cdot \vec{J_B}}|$:
\[
|\mn{\vec{J_A}\cdot \vec{J_B}}| \leq \sum_i p_i |\mn{\vec{J}}_{\rho_i}\cdot \mn{\vec{J}}_{\sigma_i}|=
\]
\[=\sum_i
 p_i |\mn{J_z}_{\rho_i'} \mn{J_z}_{\sigma_i'}|\leq \mn{N_A N_B},
\]
where the ' sign on the states $\rho_i$ and $\sigma_i$ means that we choose the basis where only the $z$ component of $\mn{\vec{J}}_{\rho_i'}$ is nonzero. Hence, for all separable states one has $W\le0$ and the positivity of $W$ witnesses the presence of entanglement.

\subsection{On entanglement in $C^2\otimes C^2$}
In the beginning of this section we showed how the expectation value of the correlation visibility can be obtained for each type of amplification. Here our perspective will be completely different: we are given a correlation visibility (that is experimentally measured) and have to decide whether it proves or not the presence of entanglement in the input state. Within this perspective the presence of entanglement is ensured by the value of the correlation visibility only if this value is impossible to obtain for a separable input state with any detection scheme.
Let us first introduce the general detection, as usual its representation in a basis $\{a, a_\perp\}$ is given by the following POVM:
\begin{eqnarray}\label{POVM}
P_+ =\eta \ket{1}\!\!\bra{1}+ \frac{1}{2}\xi \id \hspace{10 pt} P_- = \eta \ket{1_\perp}\!\!\bra{1_\perp}+\frac{1}{2}\xi \id
\nonumber\\
P_\text{null}= (1-\eta -\xi) \id \hspace{10 pt} \text{with}\hspace{10 pt}\eta + \xi < 1,
\end{eqnarray}
which describes a detector with the efficiency $\eta$ and noise rate $\xi$. The total probability of a conclusive result is then a multiple of identity $P_+ + P_- = (\eta + \xi)\id$ and $P_+ - P_- = \eta (\ket{1}\!\!\bra{1}-\ket{1_\perp}\!\!\bra{1_\perp})$. The correllation visibility contribution after the post-selection (\ref{vispost}) is given by
\begin{equation}
\frac{\eta}{\eta+\xi} \mn{(\ket{1}\!\!\bra{1}-\ket{1_\perp}\!\!\bra{1_\perp})_\textbf{a} (\ket{1}\!\!\bra{1}-\ket{1_\perp}\!\!\bra{1_\perp})_\textbf{b}}.
\end{equation}
The total visibility $V =V_{a, a_\perp}+V_{a', a'_\perp}+V_{a'', a''_\perp}$, which is the sum of the contributions of the three orthogonal bases on the Pointcar\'e sphere, is equal to:
\begin{equation}
|V| =  \frac{\eta}{\eta+\xi} |\mn{\vec{\sigma}_{\textbf{a}}\cdot \vec{\sigma}_{\textbf{b}}}|,
\end{equation}
where $\vec{\sigma}$ is a vector of Pauli matrices ($\sigma_x,\sigma_y,\sigma_z$).
The value of $|V|$ is maximal for  $\xi = 0$.
Using the entanglement witness (\ref{witness}), which for two entangled photons $\textbf{a}$ and $\textbf{b}$ reads $W = |\mn{\vec{\sigma}_{\textbf{a}}\cdot \vec{\sigma}_{\textbf{b}}}|-1$, we find that for a separable state:
\begin{equation}\label{visbound}
 |V_{a, a_\perp}+V_{a', a'_\perp}+V_{a'', a''_\perp}| \leq 1.
\end{equation}
The last inequality gives a bound above which the observed visibility proves that the input state is entangled.
We can rewrite this bound in terms of the contribution to the visibility $V_{a,a_\perp}$ from one setting $\{a,a_\perp\}$. For the the universal cloner all three orthogonal settings give the same contribution because of its symmetry and ($\ref{visbound}$) becomes \begin{equation}\label{EqBound}
|V_{a,a_\perp}|\leq 1/3.
\end{equation}
For cloning machines with equatorial symmetry there are two settings on the equator that give an equally good contribution to the visibility $V_{a, a_\perp}$~(\ref{singletVis}), and the $h$-$v$ contribution that is identically zero for the {\it measure \& prepare} cloner and negligible for the phase covariant cloner (see discussion at the end of the corresponding sections). So for these two cloning machines the bound ($\ref{visbound}$) reads
\begin{equation}\label{UBound}
|V_{a,a_\perp}|\leq 1/2.
\end{equation}

\subsection{Universal Cloner}
The evolution for the universal amplification is known to decouple for the two orthogonal polarizations modes.
For one of the polarization modes $C_A \pi_\eta(z)^{a^\dag a} C_A^\dag$ is $
\pi_\eta(z)^{a^\dag a}(C_g^2-S_g^2 \pi_\eta(z))^{-a^\dag a-1}$, and the function $\tilde{\Pi}(\eta, g)$ is a product of two such terms for both polarization modes. The restriction to the input Hilbert space gives:
\begin{align}\label{UCpi}
\tilde{\Pi}(\eta, g)_{++} &= \pi_\eta(z)(C_g^2-S_g^2 \pi_\eta(z))^{-2}(C_g^2-S_g^2 \pi_\eta(z_\perp))^{-1},
\nonumber\\
\tilde{\Pi}(\eta, g)_{--} &= (C_g^2-S_g^2 \pi_\eta(z))^{-1}\pi_\eta(z_\perp)(C_g^2-S_g^2 \pi_\eta(z_\perp))^{-2},\nonumber
\\
\tilde{\Pi}(\eta, g)_{00} &= (C_g^2-S_g^2 \pi_\eta(z))^{-1}(C_g^2-S_g^2 \pi_\eta(z_\perp))^{-1}.
\end{align}
All the non-diagonal terms are $0$. The different functions $\tilde{\Pi}_{ij}$, together with one of the three relations (\ref{generating}), allow the direct calculation of the corresponding detection probability, which can be used to find the visibility (\ref{singletVis}).

\begin{figure}[h]
\center
\includegraphics[width=8 cm]{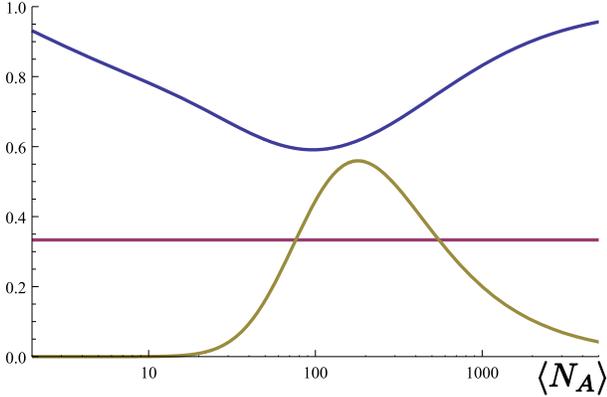}\\
  \caption{Correlation visibility $V_{a,a_\perp}$ (\ref{singletVis}) and the probability of a conclusive detection for the universal cloner
  as a function of the mean photon number $\mn{N_A}$ on the amplified side. The horizontal line is the bound 1/3 (\ref{UBound}).}
\end{figure}

\subsection{Phase Covariant Cloner}

\begin{figure}[h]
\center
\includegraphics[width=8 cm]{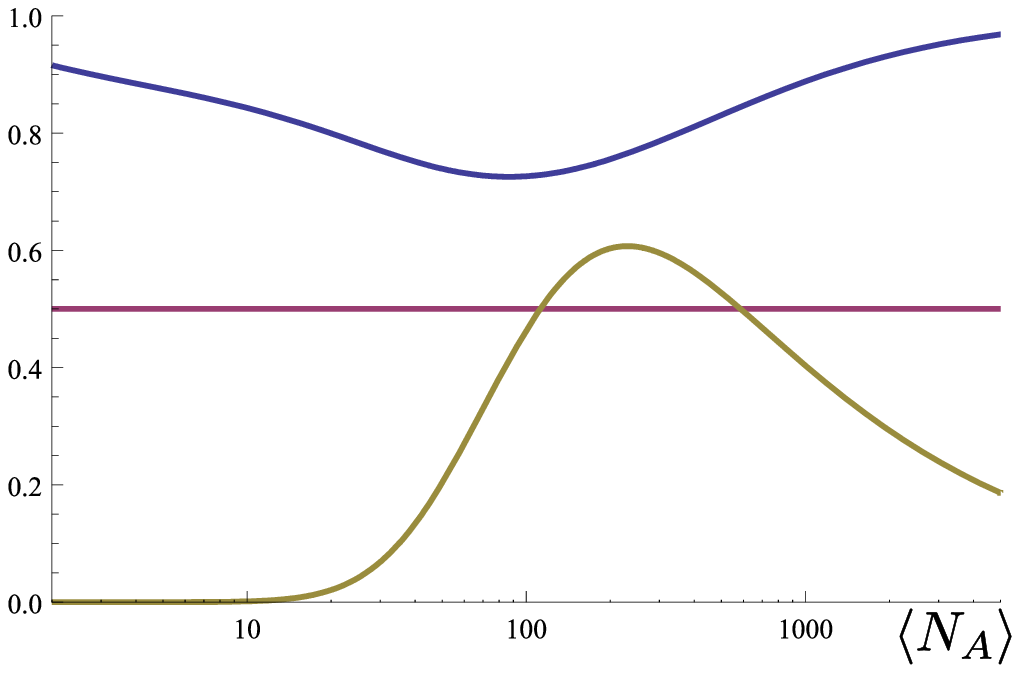}\\
  \caption{Correlation visibility $V_{a,a_\perp}$ (\ref{singletVis}) and the probability of a conclusive detection for the phase covariant cloner
  as a function of the mean photon number $\mn{N_A}$ on the amplified side. The horizontal line is the bound 1/2 (\ref{EqBound}).}
\label{fig:PCCV}
\end{figure}

The situation is similar for the Phase Covariant Cloner. The evolution is decoupled for the two polarization modes, the generating function is transformed to $U^\dag  \pi_\eta(z)^{a^\dag a} U$, that is the complicated expression (\ref{Pccgen}), where we replace $e^{k} \rightarrow \pi_\eta(z)$. But when we make the restriction we obtain a simpler form that is similar to (\ref{UCpi}):
\begin{align}
\tilde{\Pi}(\eta, g)_{++} = \frac{\pi_\eta(z)}{ (C_g^2-S_g^2 \pi_\eta(z)^2)^{-\frac{3}{2}} (C_g^2-S_g^2 \pi_\eta(z_\perp)^2)^{-\frac{1}{2}}},
\nonumber \\
\tilde{\Pi}(\eta, g)_{--} =  \frac{\pi_\eta(z_\perp)}{(C_g^2-S_g^2 \pi_\eta(z)^2)^{-\frac{1}{2}}  (C_g^2-S_g^2 \pi_\eta(z_\perp)^2)^{-\frac{3}{2}}},\nonumber
\\
\tilde{\Pi}(\eta, g)_{00} =  (C_g^2-S_g^2 \pi_\eta(z)^2)^{-\frac{1}{2}}(C_g^2-S_g^2 \pi_\eta(z_\perp)^2)^{-\frac{1}{2}}.
\end{align}
The values of the visibility for different gains are given in the Fig.~\ref{fig:PCCV}.

\subsection{Measure \& Prepare}

The evolution of the input state $\rho_0$ through the {\it measure \& prepare} amplifier is given in (\ref{mp}), making it act on $\pi_\eta(z)^{a^\dag a} \pi_\eta(z_\perp)^{a_\perp^\dag a_\perp}$ gives:
\begin{equation}
\int \frac{d\phi}{\pi} \ket{1_\phi}\!\!\bra{1_\phi} \bra{\alpha c_\phi}\pi_\eta(z)^{a^\dag a}\ket{\alpha c_\phi}\bra{\alpha s_\phi}\pi_\eta(z_\perp)^{a_\perp^\dag a_\perp}\ket{\alpha s_\phi},
\end{equation}
that is :
\[\tilde{\Pi}(\eta \alpha)=\int \frac{d\phi}{\pi}  e^{-|\alpha\eta|^2(1- z c_\phi^2- z_\perp s_\phi^2)} \ket{1_\phi}\!\!\bra{1_\phi}. \]
To make the restriction to the initial Hilbert space we simply rewrite $\ket{1_\phi}\!\!\bra{1_\phi}$ in the basis $\{\ket{1},\ket{1_\perp}\}$:
\begin{equation}
\tilde{\Pi}(\eta\alpha)|_{H_i}=\int \frac{d\phi}{\pi}
\left (
\begin{array}{ccc}
c_\phi^2 & c_\phi s_\phi & 0 \\
c_\phi s_\phi & s_\phi^2 & 0 \\
0&0& 0
\end{array}
\right )e^{-|\alpha\eta|^2(1- z c_\phi^2- z_\perp s_\phi^2)}.
\end{equation}
The integration kills the off diagonal elements while the diagonal can be expressed with the modified Bessel function $I_0(z)=\int \frac{d\phi}{2\pi} e^{- z c_\phi} $ as:
\begin{align}
\tilde{\Pi}(\eta\alpha)_{++} &= 2 \frac{e^{-|\eta\alpha|^2}}{|\eta\alpha|^2}\partial_z (e^{|\eta\alpha|^2\frac{z+z_\perp}{2}} I_0(  |\eta\alpha|^2\frac{z-z_\perp}{2})),
\nonumber \\
\tilde{\Pi}(\eta\alpha)_{--}&=2\frac{e^{-|\eta\alpha|^2}}{|\eta\alpha|^2}\partial_{z_\perp} ((e^{|\eta\alpha|^2\frac{z+z_\perp}{2}} I_0(  |\eta\alpha|^2\frac{z-z_\perp}{2})).
\end{align}
\begin{figure}[h]
\center
\includegraphics[width=8 cm]{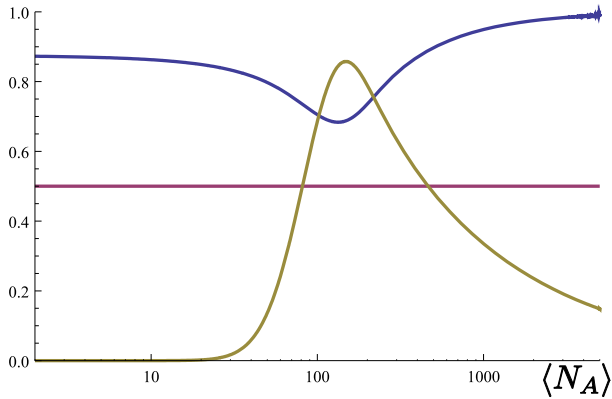}\\
  \caption{Correlation visibility $V_{a,a_\perp}$ (\ref{singletVis}) and the probability of a conclusive detection
   as a function of the mean photon number $\mn{N_A}$ on the amplified side for the {\it measure \& prepare} cloner. The horizontal line is the bound 1/2 (\ref{EqBound}). The curves in this figure are sharper with respect to the two previous cases because the photon number distribution for a coherent state (Poissonian) is narrower than the distributions of states produced by the phase covariant and the universal cloners (Super-poissonian).}
\end{figure}
\subsection{What does the correlation visibility tell us?}
In Fig.~4-6 we plotted the observed correlation visibility $V_{a,a_\perp}$, the probability of a conclusive event $P$ (which tells us the amount of post-selection) and the bound above which the visibility proves the entanglement of the initial photon pair. We see that with each of the three different amplification types we achieve visibilities that are higher than this bound, consequently one can ``see'' micro-micro entanglement with any of the cloning machines described. Surprisingly the highest observed visibility is that seen with the {\it measure \& prepare} cloner, and we know that its output is not entangled with Bob's photon (the entanglement is broken by the projective measurement done by the cloner). This illustrates the fact that the measurement of correlation visibility
does not prove the micro-macro entanglement, indeed the separable state prepared by the {\it measure \& prepare} cloner shows that such a naive analysis doesn't apply.

We can also analyze the observed correlation visibility from the perspective of Bell inequalities. The famous CHSH inequality introduced in \cite{CHSH} gives a bound on the observed visibility:
\begin{equation}\label{chshin}
|V^\text{CHSH}_{a,a_\perp}| >\frac{1}{\sqrt{2}}\approx 0.7,
\end{equation}
below which the correlations can not be reproduced by any local model. As shown in Fig.~6, the correlation visibility observed for the {\it measure \& prepare} cloner can be higher then the CHSH bound. A priori this seems to be a paradox, however it is well known that the post-selection opens the detection loophole, in particular in the analysis of Clauser {\it et al.}~\cite{CHSH} no measurement result can be disregarded. Again the importance of the detection loophole is fully revealed by this example. Remark that without post-selection the observed correlation visibility is given by $P V_{a,a_\perp}$, which never exceeds the CHSH bound $1/\sqrt{2}$ for all of the presented cloning machines.

To finish we note that these results can be generalized for other types of threshold detectors, such as the {\it  orthogonality filter} introduced in \cite{Rome2}.

%
%

\section{Micro-macro entanglement}\label{entanglement}
In this section we ask ourselves whether the different cloning machines presented in section \ref{cloning} preserve entanglement, that is whether after amplification of Alice's photon, Bob's single-photon is entangled with Alice's many photons. Here we consider this as a question of principle, hence we do not consider which measurement could reveal the entanglement.

To apply the criteria (\ref{witness}) to the present case we choose a basis for the vectors, let it be the poles of the Bloch sphere - $h$-$v$ component denoted ``$z$'' and two components on the equator denoted ``$x$'' and ``$y$''. Each of the components $J^A_i J^B_i= J_i \sigma_i$ in this basis has the form $(a_i^\dag a_i -a_{i\perp}^\dag a_{i\perp})(b_i^\dag b_i - b_{i\perp}^\dag b_{i \perp})$. For this purpose we place ourselves in the Heisenberg representation and apply the evolution on these operators while the state remains $\ket{\psi^-}$. The $b$ photon operators are not affected by the evolution, all that can happen is that we lose Bob's photon, but as we use it to trigger the experiment these cases are excluded (for the same reason $N_B= 1$). Thus we know the value of $b^\dag b$ at all times and can make the average on Bob's side so that:
\begin{equation}
\bra{\psi^-} J_i^t\sigma_i^t \ket{\psi^-} = \bra{1_i}(a_i^\dag a_i)_t-(a_{i\perp}^\dag a_{i\perp})_t\ket{1_i},
\end{equation}
where  $\ket{1_i}= a_i^\dag \ket{0}$. Remark that a transmission loss after the amplification has a simple effect $W \to \eta W$ with $\eta$ - the transmission efficiency.

\subsection{Phase Covariant Cloner}

\begin{figure}[h]
\center
  \includegraphics[width=8 cm]{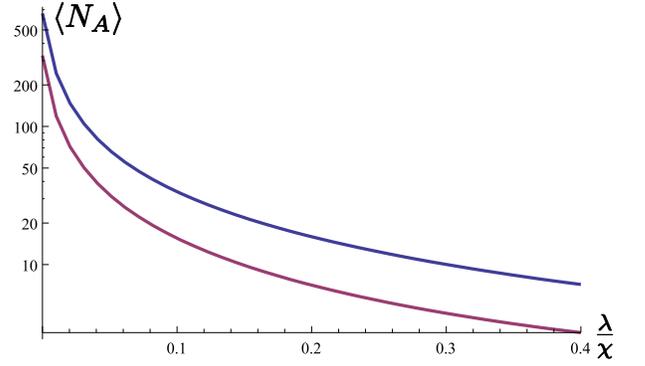}\\
  \caption{Number of photons in the amplified state for which the violation is $0$ and $1$ as a  function of the ratio $\frac{\lambda}{\chi}$ for the phase covariant cloner.}
  \label{fig:mmpcc}
\end{figure}

In the ideal case the phase covariant amplification leads to a unitary evolution, in which case the initial entanglement is trivially preserved. However if there are some losses anywhere the unitarity is broken. Here we treat the case where losses occur during cloning. For the phase covariant amplification we know that $J_x \sigma_x = J_y \sigma_y$ and because the evolution of $a$ and $a_\perp$ decouples $\mn{J_y\sigma_y} = \bra{1_y}(a_y^\dag a_y)_t\ket{1_y}- \bra{0}(a_y^\dag a_y)_t\ket{0}$ while the total photon number expressed in an equatorial basis is $ N = \bra{1_y}(a_y^\dag a_y)_t\ket{1_y}+ \bra{0}(a_y^\dag a_y)_t\ket{0}$ implying that:
\begin{eqnarray}
W = \bra{1_y}(a_y^\dag a_y)_t\ket{1_y} - 3\bra{0}(a_y^\dag a_y)_t\ket{0}+ \mn{J_z \sigma_z}.
\end{eqnarray}
The operator $(a_i^\dag a_i)_t$ obeys the damped Heisenberg equation:
\begin{equation}\label{dumped}
\frac{d}{dt}(a_i^\dag a_i)_t = i [H ,(a_i^\dag a_i)_t] - \lambda (a_i^\dag a_i)_t -\hat{f},
\end{equation}
where the damping is a result of a weak interaction with a collection of empty environmental modes (such as absorption and geometrical scattering), $\hat{f}$ is the Langevin-noise fluctuation induced by the statistics of these modes. We take $\mn{\hat{f}} = 0$ which corresponds to the environment in the vacuum state before the interaction. In the equatorial modes the evolution is diagonal for the quadrature operators $x=\frac{1}{\sqrt{2}}(a_y^\dag + a_y)$ and $p=\frac{i}{\sqrt{2}}(a_y^\dag - a_y)$. With $(a_y^\dag a_y)_t= \frac{1}{2}(x_t^2+p_t^2-1)$ and $H=\frac{\chi}{2}(xp+px)$ the equations of motion for the squares of the quadratures are:
\[
\left\{
  \begin{array}{ll}
    \frac{d}{dt} \mn{x_t^2}=(2\chi -\lambda) \mn{x_t^2}+\frac{1}{2}\lambda ,\\
    \frac{d}{dt} \mn{p_t^2}=(-2\chi -\lambda) \mn{p_t^2}+\frac{1}{2}\lambda,
  \end{array}
\right.
\]
which is straightforward to solve but yields a long expression. In the polar modes ``z'' the commutator $[H, a^\dag_h a_h - a^\dag_v a_v] = 0$ and the only thing that acts on this operator is damping, thus $\mn{J_z \sigma_z} = e^{-\lambda t}$. After rescaling  $(\lambda t ,\chi t) \to (\lambda,\chi) $ we obtain:
\begin{equation}
W= 1+e^{-\lambda } - \frac{2\lambda}{(2\chi)^2-\lambda^2}(e^{-\lambda }(2 \chi S_{2 \chi }+\lambda C_{2 \chi })-\lambda),
\end{equation}
with $S_{2 \chi }= \sinh {2 \chi} $ and $C_{2 \chi }=\cosh{2 \chi }$. While the number of photons is:
\begin{equation}
N=
2e^{-\lambda} C_{2\chi}+
\frac{2\lambda(e^{-\lambda }(2 \chi S_{2 \chi } +\lambda C_{2 \chi })-\lambda)}{(2\chi)^2-\lambda^2}.
\end{equation}
In Fig.~\ref{fig:mmpcc} we plot the parametric curves for $W=1$ and $W=0$ as functions of the total photon number and the amplification/damping ratio that describes the quality of the amplification. Below the top curve we are sure to find some persistent entanglement in the system.

\subsection{Universal Cloner}

\begin{figure}[h]
\center
  \includegraphics[width=8 cm]{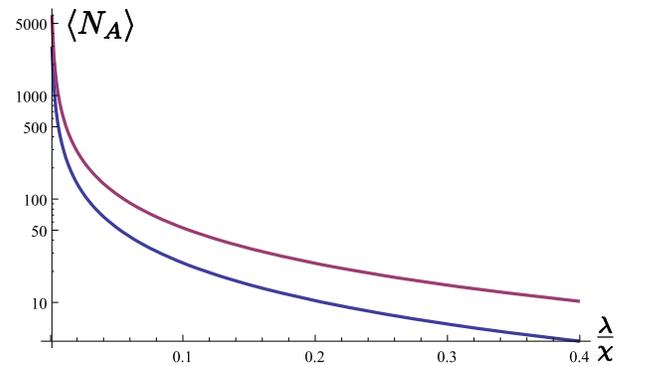}\\
  \caption{Number of photons in the amplified state for which the violation is $0$ and $1$ as a function of the ratio $\frac{\lambda}{\chi}$ for the universal cloner.}
  \label{fig:mmuc}
\end{figure}

For the universal cloner all the three components of $\mn{\vec{J}\cdot\vec{\sigma}}$ are the same, and with the photon number being $N=\bra{1}(a^\dag a)_t\ket{1}+ \bra{0}(a^\dag a)_t\ket{0}$ the violation becomes $W= 2\bra{1}(a^\dag a)_t\ket{1}- 4 \bra{0}(a^\dag a)_t\ket{0}$ for an arbitrary mode, let it be the mode $a_h$. The damped Heisenberg equation~(\ref{dumped}) for the Hamiltonian (\ref{universal}) gives the complete set of differential equations:
\[
\left\{
    \begin{array}{ll}
    \partial_t
\mn{a_h^{\dagger }a_h} =
\text{ $\chi $ }
\left(\mn{a_h^{\dagger }c_v^{\dagger }}+\mn{a_h c_v}\right.
\text{) - $\lambda $ }
\mn{a_h^{\dagger }a_h}, \\
\partial _t
\mn{c_v^{\dagger }c_v} =
\text{ $\chi $ }
\left(\mn{a_h^{\dagger }c_v^{\dagger }}+\mn{a_h c_v}\right.
\text{) - $\lambda $ }
\mn{c_v^{\dagger }c_v},\\
\partial _t
\mn{a_h^{\dagger }c_v^{\dagger }} =
\text{ $\chi $ }
\left(\mn{a_h^{\dagger }a_h}+ \mn{c_v^{\dagger }c_v} + 1 \right.
\text{) - $\lambda $ }
\mn{a_h^{\dagger }c_v^{\dagger }},\\
\partial _t
\mn{a_h c_v} =
\text{ $\chi $ }
\left(\mn{a_h^{\dagger }a_h}+ \mn{c_v^{\dagger }c_v} + 1 \right.
\text{) - $\lambda $ }
\mn{a_h c_v}.
\end{array}
\right.
\]
The above system can be diagonalized and solved by standard linear algebra methods yielding (for $\mn{c_v^\dag c_v}_0 = 0$):
\begin{align}
W &= 2 G(\chi,\lambda) - 2Q(\chi,\lambda), \\
N &=  G(\chi,\lambda) + 2Q(\chi,\lambda),
\end{align}
with $G(\chi,\lambda)= \frac{1}{2}e^{-\lambda}(\text{Cosh}(2\chi  )+1)$ and
$ Q(\chi,\lambda)= \frac{1}{4}\left(\frac{2\chi }{2\chi -\lambda }\left(e^{2\chi -\lambda }-1\right)+\frac{2\chi }{2\chi +\lambda }\left(e^{-2\chi -\lambda }-1\right)\right)$. In the Fig.~\ref{fig:mmuc} we plot the curves $W=1$ and $W=0$ in the plain defined by the number of photons and the quality of the amplification.

\subsection{Loss Before The Amplification}

The entanglement witness $W$ is very fragile in the case where the singlet state is subject to loss of the photon $\textbf{a}$ in the amplified spatial mode (the presence of the other photon $\textbf{b}$ is ensured by the heralding). The input state in this case is:
\begin{equation}
\rho = p \ket{\psi^-}\!\!\bra{\psi^-} + \frac{1-p}{2}(b^\dag \ket{0}\!\!\bra{0} b+b^\dag_\perp \ket{0}\!\!\bra{0}b_\perp).
\end{equation}
The value of $W$ is then the sum of the two contributions for the components $p \, \rho_e =p \ket{\psi^-}\!\!\bra{\psi^-}$ and $(1-p) \rho_s = \frac{1-p}{2}(b^\dag \ket{0}\!\!\bra{0} b+b^\dag_\perp\ket{0}\!\!\bra{0}b_\perp)$. The first one gives the violation for the ideal lossless case for both the universal cloner and the phase covariant cloner $p \mn{W}_{\rho_e} = 2 p$. The second term however becomes negative very fast because $\mn{\vec{J}_i}$ is $0$ for the amplified vacuum $(1-p) \mn{W}_{\rho_s} =- (1-p) \mn{N}_{\rho_s} = -2(1-p)n_0$, where $n_0$ is the number of photons for the amplified vacuum. Consequently $W$ is positive iff:
\begin{equation}
n_0 \leq \frac{p}{1-p}.
\end{equation}
In order to prove the entanglement with this witness we must impose a condition on $p$ which becomes more and more restrictive when the number of photons in the amplified state grows. The general situation is however different for the universal and the phase covariant amplifiers. For the universal cloner we can indeed show that $\$_A(\rho)$ becomes separable for $n_0 = \frac{p}{1-p}$. While under the assumption of unitarity , the phase covariant cloner preserves the entanglement for any $p \neq 0$ and $n_0$, this is the simple consequence of the fact that the two states $\rho_e$ and $\rho_s$ live in independent sectors of the Hilbert space and consequently the state $\rho$ is always entangled (for example one can use the PPT criterion \cite{Peres}, $p \,\rho_e^{T_A}$ will always have negative eigenvalues that would also be eigenvalues of $\rho^{T_A}$) and a unitary evolution will only modify the eigenvectors but not the eigenvalues, note that this simple image doesn't work when we add loss after the amplification.

\section{Conclusion}

In conclusion, if one clones one photon from an entangled pair of photons, then one can {\it see} quantum correlations with the naked eye. However, the observed correlation allow one to conclude only about entanglement of the initial 2-photon pair: one can {\it see} only micro-micro entanglement. Nevertheless, cloning with amplifiers - in contrast to measure~\&~prepare cloners - does preserve some entanglement; hence there is some micro-macro entanglement, though the detection of micro-macro entanglement requires more sophisticated detection schemes than mere naked eyes.

In this article we also emphasized the danger of eliminating inclusive data. In entanglement studies this is an acceptable procedure, as one works within standard Hilbert space quantum physics and can thus trust one's detectors and theory thereof. For the study of nonlocality, however, eliminating inconclusive results can easily lead to wrong conclusion. This illustrates on a practical example the importance of the detection loophole.

Several questions are left open. For example, one could consider cloning both photons of the initial pair. Another natural question concerns the micro-macro entanglement: how much entanglement is there? Can one detect it experimentally with realistic detectors? We leave these fascinating questions for future work. Likewise, the experimental realizations of the analyzed situations, using the various cloners, are underway and will be discussed in future works.

\section{Acknowledgements}

We would like to thank our colleagues Hugo Zbinden, Nicolas Brunner  and Cyril Branciard for useful discussion and comments.
This work was supported by the Swiss FNRS.

\appendix
\section{Spin groups}\label{algebra}

Let us remind the following algebraic relation, which is very useful in the context of continuous groups

\begin{equation}
e^B A e^{-B} = \sum_{n\geq 0} \frac{1}{n!}[B,A]^{(n)}
\end{equation}

where by $[B,A]^{(n)}$ we mean the nested commutator $[B,\dots [B,A]\dots]$ with $B$ appearing $n$ times. The equality can be easily proven by repeatedly differentiating (expanding in its Taylor series) the expression $e^{\lambda B} A e^{- \lambda B}$ and putting $\lambda = 1$ at the end. Trivially
\begin{equation}
e^B f(A) e^{-B} = f(\sum_{n\geq 0} \frac{1}{n!}[B,A]^{(n)})
\end{equation}
for any analytical function $f$.
Now, following \cite{Collett} we introduce a set of operators $\sigma_+, \sigma_-, \sigma_z$ with spin commutators
\[
[\sigma_+,\sigma_-]=\sigma_z \hspace{40 pt} [\sigma_z,\sigma_\pm]=\pm 2 \sigma_\pm
\]
In the following we will consider hamiltonians that are time-dependent linear combinations of these operators. Applying (2) we derive some formulas useful for rearranging propagators.
\begin{eqnarray}
e^{k \sigma_z} f(\sigma_\pm)= f(e^{\pm 2k}\sigma_\pm)e^{k \sigma_z}, \\
e^{k \sigma_\pm} f(\sigma_z)= f(\sigma_z \mp 2k \sigma_\pm)e^{k \sigma_\pm}, \\
e^{k \sigma_\mp} f(\sigma_\pm)= f(\sigma_\pm \mp k \sigma_z - k^2 \sigma_\mp)e^{k \sigma_\mp}.
\end{eqnarray}
Adding some structure we can identify two different sets $\{\sigma_+, \sigma_-, \sigma_z \}_\pm$ according to  $\sigma_+^\dag = \pm \sigma_-$. In both cases we construct unitary operators of the form
\[
U_+ = e^{\lambda(\sigma_+-\sigma_-)}\qquad U_-=e^{\lambda(\sigma_++\sigma_-)}.
\]
We rewrite these expressions as:
\begin{eqnarray}\label{desentanglement}
U_+=e^{\tan \lambda \,\sigma_+}e^{-\ln ( \cos \lambda )\sigma_z} e^{-\tan \lambda \,\sigma_-},
\nonumber\\
U_-=e^{\tanh \lambda \,\sigma_+}e^{-\ln (\cosh \lambda)\sigma_z} e^{\tanh \lambda \,\sigma_-}.
\end{eqnarray}
This can be shown as follows. Suppose $U_{\pm} = e^{f_\lambda\sigma_+}e^{g_\lambda\sigma_z}e^{h_\lambda\sigma_-}$ derivate both sides by $\lambda$, then pull all the pre-exponential factors on the left using (3)-(5) and multiply by $U_\pm^\dag$ from the right. We obtain
\[
\sigma_+ \pm \sigma_- =
f' \sigma_+ +g'(\sigma_z-2f\sigma_+)+h' e^{-2g_\lambda}(\sigma_-+f\sigma_z-f^2\sigma_+),
\]
that yields a system of differential equations, solved to (6),(7). In a similar way one can prove
\begin{eqnarray}
e^{\lambda \sigma_\pm}e^{\mu\sigma_\mp}=
e^{\frac{\mu\sigma_{\mp}}{1+\mu\lambda}}
e^{\pm\ln (1+\mu\lambda)\sigma_z}
e^{\frac{\lambda\sigma_{\pm}}{1+\mu\lambda}},\\
e^{\lambda \sigma_z +\mu\sigma_\pm}=
e^{ \mu \frac{e^{\pm 2 \lambda}-1}{\pm 2\lambda} \sigma_\pm}
e^{\lambda \sigma_z}.
\end{eqnarray}
The difference between the two sets appears clearly, the propagator $U_+$ is compact while $U_-$ is non-compact that will correspond to very different physical situations. Now we give some example of particular realizations of these systems
\begin{center}
\begin{tabular}[b]{c @{\hspace{10pt}}c @{\hspace{10pt}} c}
$\sigma_+$ & $\sigma_-$ & $\sigma_z$ \\
\hline
$a\, b^\dag$ & $a^\dag b$& $ b^\dag b - a^\dag a$ \\
$\frac{1}{2}{a^\dag}^2$ &  $-\frac{1}{2}{a}^2$ & $a^\dag a + \frac{1}{2}$\\
$a^\dag b^\dag$ & $- a\, b$ & $1+a^\dag a+ b^\dag b$\\
\end{tabular}
\end{center}
The first case is the Schwinger representation of the spin group, that describes the interaction of modes $a$ and $b$ on a beam splitter and belongs to the compact $U_+$ class. The second and the third cases are in the $U_-$ class and correspond to the spontaneous emission driven by a classical pump laser.

\section{Projectors in Fock space}\label{Fock space projections}

Given a quantum state $\rho = \sum \rho_{nm}\ket{n}\!\!\bra{m}$ with the mode $a$ expressed in the Fock space components, we introduce the operators $\hat{\Pi}(z)= z^{a^\dag a}$ and $\hat{N}(k)= e^{k a^\dag a}$ with resulting characteristic functions
\begin{eqnarray}
\Pi(z)= \textrm{tr}\, \rho \,z^{a^\dag a},
\\
N(k)=\textrm{tr}\, \rho \,e^{k a^\dag a},
\end{eqnarray}
that are related via $N(k)=\Pi(e^{k})$. The first function is useful to derive projections of the state on Fock components, that can describe the perfect detection process
\begin{eqnarray}
\frac{1}{n!} \partial_z^n \Pi(z)|_{z=0}= \rho_{nn},
\\
\frac{1}{n!}\partial_z^n\frac{\Pi(z)}{1-z}|_{z=0}=\sum_{k\geq0}^n\rho_{kk},
\end{eqnarray}
that are projections of the the state on $\ket{n}\!\!\bra{n}$ and $\sum_k^n\ket{k}\!\!\bra{k}$ respectively. The second
function $N(k)$ gives mean values of powers of the number of particles operator
\begin{equation}\label{Nfunction}
\partial_k^n N(k)|_{k=0}= \textrm{tr}\, \rho \,(a^\dag a)^n.
\end{equation}
It also can be used used to express higher order correlation functions through $e^{k a^\dag a}=:e^{(e^{k}-1) a^\dag a} :$. Given one of these scalar functions one can extract all the information about the diagonal terms of the density matrix.

\section{Array of single-photon detectors}\label{detectors array}
Consider an array of $N$ ideal single-photon detectors, if $N$ is a power of two this may be realized with some $50\%$ beam splitters as shown in Figure 1.
\begin{figure}[b]
\center
\includegraphics[width=5 cm]{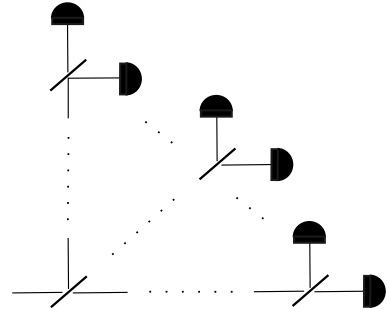}\\
  \caption{An array of detectors $N=n^2$, realized with $n$ beam splitters.}
\end{figure}
However this restriction is not necessary, all we need is all detectors to be the same. Now one may define a "see" event by all events when at least $\theta$ detectors click. The other "not see" case happens when only less than $\theta$ detectors click is given by the operator
\begin{equation}
\hat{P_{ns}^\theta} = \sum_{j \geq 0}^{\theta} \hat{P_{j}},
\end{equation}
where $\hat{P_{j}}$ stands for an event where exactly $j$ detectors click and it is equal to
\begin{equation}\label{PJ}
\hat{P_{j}}= C_j^N (1-\prjct{0})^{\otimes j}\otimes \prjct{0}^{\otimes(N-j)},
\end{equation}
where $C_j^N$ corresponds to the number of ways of picking $j$ detectors out of $N$. Expand the last expression
\[
\hat{P_j} = \sum_{k\geq 0}^j C_k^j C_j^N (-1)^{j-k}\prjct{0}^{\otimes N-k}.
\]
The projector $\prjct{0}^{\otimes N-k}$ describes the event when $N-k$ detectors don't get any photons thus all the $a^\dag a$ photons go into $k$ other detectors, the probability of such an event is $(\frac{k}{N})^{a^\dag a}$. If now we take the limit $N \rightarrow \infty$ and apply the Stirling approximation to $\frac{N!}{(N-j)!}$, (\ref{PJ}) becomes
\begin{equation}
\hat{P_j} =  \lim_{N\rightarrow \infty} \sum_{k\geq 0}^j \frac{(-1)^{j-k}k^{a\dag a}}{j!(j-k)!} N^{j-a\dag a}.
\end{equation}
This expression is obviously zero for $a^\dag a > j$, while in the other case we will show in appendix that
\begin{equation}
\sum_{k\geq 0}^j \frac{(-1)^{j-k}k^{a\dag a}}{j!(j-k)!} = \delta_{a^\dag a}^j \qquad |\quad a^\dag a \leq j,
\end{equation}
yielding $\lim_{N \rightarrow \infty } \hat{P_j} = \delta_{a^\dag a}^j$ that implies in the limit $N$ goes to infinity
\begin{equation}
\hat{P_{ns}^\theta}= \sum_{j\geq 0}^\theta \delta_{a^\dag a}^j = \sum_{j\geq 0}^\theta \prjct{j},
\end{equation}
to account for a finite efficiency of detectors we add a loss as in previous section.

\section{Universal cloning with doped fibers}\label{fibers}

Suppose a dipole (two level system) oriented along a certain direction and coupled resonantly to an optical mode $a_\beta$, with the polarization being the projection of the dipole orientation on the plane defined by the $k$-vector. The atom is in the excited state $\ket{e}$, as usual one introduce the transition operators $\sigma_+ = \ket{e}\!\!\bra{g}$ and $\sigma_- = \ket{g}\!\!\bra{e}$. The interaction hamiltonian in the rotating wave approximation is $ H_i = \kappa i( a_\beta^\dag \sigma_- - a_\beta \sigma_+) $ with the propagator being $U_\beta= \exp(-i H_i t)$, because we are dealing with a two level system with $\sigma_-^2 = \sigma_+^2 = 0$ and
\[
(a_\beta^\dag \sigma_- - a_\beta \sigma_+)^{2n}= (-1)^n ((a^\dag a)^n \ket{g}\!\!\bra{g}+(a a^\dag )^n \ket{e}\!\!\bra{e}),
\]
\[
(a_\beta^\dag \sigma_- - a_\beta \sigma_+)^{2n+1}= (-1)^n ((a^\dag a)^n a^\dag \sigma_+ -(a a^\dag )^n a \sigma_-),
\]
the exponential can be expanded and resummed yielding
\[
U_\beta = \cos(\kappa t \sqrt{a_\beta^\dag a_\beta}) \ket{g}\!\!\bra{g} +\cos(\kappa t \sqrt{a_\beta a_\beta^\dag }) \ket{e}\!\!\bra{e} +
\]
\[
+\frac{\sin(\kappa t \sqrt{a_\beta^\dag a_\beta})}{\sqrt{a_\beta^\dag a_\beta}} a_\beta^\dag \sigma_- - \frac{\sin(\kappa t \sqrt{a_\beta a_\beta^\dag })}{\sqrt{ a_\beta a_\beta^\dag }} a_\beta \sigma_+.
\]
The interaction of a single atom with a propagating optical mode can be reasonable considered small to the first order, in which case the evolution becomes
\[
U_\beta = (1- \delta g \frac{1}{2} a_\beta^\dag a_\beta) \ket{g}\!\!\bra{g} +(1-\delta g \frac{1}{2} a_\beta a_\beta^\dag ) \ket{e}\!\!\bra{e} +
\]
\[
+\sqrt{\delta g} a_\beta^\dag \sigma_- - \sqrt{\delta g} a_\beta \sigma_+,
\]
with $\delta g$ - the small interaction parameter. If in the neighborhood of the dipole "$\beta$" there is a dipole oriented in an orthogonal way " $\beta_\perp$" that couples to the orthogonal polarization, we should add the term $U_{\beta_\perp}$ to the evolution operator. Finally after tracing over the atoms in the exited states the optical state becomes
\[
\rho(g+\delta g) = \text{tr}_{\beta,\beta_\perp} U_\beta U_{\beta_\perp} \rho(g)\otimes\ket{e}\!\!\bra{e}\otimes \ket{e}\!\!\bra{e} U_\beta^\dag U_{\beta_\perp}^\dag,
\]
in the limit $\delta g \to 0$ that defines a differential equation
\[
\dot{\rho} = -\frac{1}{2}(a a ^\dag \rho + \rho a a^\dag - 2 a^\dag \rho a +
\]
\[
+a_{\perp} a_{\perp}^\dag \rho + \rho a_{\perp} a_{\perp}^\dag - 2 a_{\perp}^\dag \rho a_{\perp}).
\]
Because this expression is invariant we did not write $\beta$, we recovered the universal symmetry. One can easily check that the solution has the form $\$_A (\rho_0)$ defined in (\ref{amplichannel}).


\begin{thebibliography}{99}

\bibitem{Sekatski} P. Sekatski, N. Brunner, C. Branciard, N. Gisin and C. Simon, Phys. Rev. Lett. {\bf103}, 113601 (2009)

\bibitem{Herbert} N. Herbert, Found. Phys. {\bf 12}, 1171 (1982).

\bibitem{WoottersZurek} W. Wootters and W.H. Zurek, Nature (London) {\bf 299}, 802 (1982).

\bibitem{Diekes} D. Dieks, Phys. Lett. A {\bf 92}, 271 (1982).

\bibitem{Milonni} P.W. Milonni and M.L. Hardies, Phys. Lett. A {\bf92},321 (1982).

\bibitem{Gisin} N. Gisin, Phys. Lett. A {\bf 242}, 1 (1998).

\bibitem{BuzekGisinSimon} V. Buzek, N. Gisin and C. Simon, Phys. Rev. Lett. {\bf87}, 170405 (2001).

\bibitem{GisinMassar} N. Gisin and S. Massar, Phys. Rev. Lett. {\bf79}, 2153 (1997).

\bibitem{Fasel} S. Fasel et al., Phys. Rev. Lett. {\bf89}, 107901 (2002).

\bibitem{Rome} E. Nagali et al., Phys. Rev. A {\bf76}, 042126 (2007).

\bibitem{Rome2}F. De Martini et al., Phys. Rev. Lett. {\bf 100}, 253601 (2008).

\bibitem{OptClonLett} C. Simon, G. Weihs, and A. Zeilinger Phys. Rev. Lett. {\bf 84}, 13 (2000).

\bibitem{OptClon} J. Kempe, Ch. Simon and G. Weihs Phys. Rev. A {\bf 62}, 032302 (2000).

\bibitem{Witness} C. Simon and D. Bouwmeester, Phys. Rev. Lett. {\bf91}, 053601 (2003).

\bibitem{optFidelities} V. Scarani et al., Rev. Mod. Phys. {\bf 77}, 1225-1256, (2005).

\bibitem{riekebaylor} F. Rieke and D.A. Baylor, Rev. Mod. Phys. {\bf70}, 1027-1036, (1998).

\bibitem{Acin} J. Bae and A. Ac\'in, Phys. Rev. Lett. {\bf97}, 030402 (2006).

\bibitem{MantisSchrimps} N.W. Roberts et al., Nature Photonics {\bf3}, 641 (2009).

\bibitem{dimH} A. Ac\'in, N. Gisin and L. Masanes, Phys. Rev. Lett. {\bf97}, 120405 (2006).

\bibitem{Collett} M. J. Collett, Phys. Rev. A {\bf38}, 2233 (1988).

\bibitem{Barlow} H. Barlow J. Opt. Soc. Am.  {\bf46}, 8 (1956).

\bibitem{Eye} S. Hecht, S. Shlaer, and M. Pirenne, J. Gen. Physiol. {\bf25}, 819 (1942).

\bibitem{CHSH} J. F. Clauser et al. Phys. Rev. Lett. {\bf 23}, 880-884, (1969).

\bibitem{Peres} A. Peres Phys. Rev. Lett. {\bf77}, 8 (1996).


\end{thebibliography}
\end{document}